\documentclass[prd,a4paper, nofootinbib, twocolumn]{revtex4}
\usepackage{graphicx}

\newcommand{\beq}{\begin{equation}}
\newcommand{\eeq}{\end{equation}}
\newcommand{\beqs}{\begin{eqnarray}}
\newcommand{\eeqs}{\end{eqnarray}}

\newcommand{\Tr}{{\rm Tr}}

\renewcommand{\L}{{\cal L}}

\newcommand{\C}{{\cal C}}
\newcommand{\M}{{\cal M}}

\renewcommand{\O}{{\cal O}}

\def\hbar{\hspace{0pt}\raisebox{1pt}{$-$} \hspace{-7pt} h}

\begin{document}
\title{Breaking Discrete Symmetries in Broken Gauge Theories}

\author{Thomas Appelquist$^{}$ \thanks{email: thomas.appelquist@yale.edu}}
\author{Yang Bai$^{}$ \thanks{email: yang.bai@yale.edu}}
\author{Maurizio Piai$^{}$ \thanks{email: maurizio.piai@yale.edu}}

\vspace{6mm}

\affiliation{\ Physics Department, Sloane Laboratory \\
Yale University \\
New Haven, CT 06520}

\begin{abstract}
We study the spontaneous breaking of discrete symmetries in
theories with broken gauge symmetry. The intended application is
to CP breaking in theories with gauged flavor symmetries, but the
analysis described here is preliminary. We dispense with matter
fields and take the gauge theory to be weakly coupled and broken
spontaneously by unspecified, short-distance forces. We develop an
effective-field-theory description of the resultant low energy
theory, and ask whether this theory by itself can describe the
subsequent breaking of discrete symmetries. We conclude that this
can happen depending on the parameters of the effective theory,
and that the intrinsic violation is naturally of order unity.
\end{abstract}

\maketitle

\section{Introduction}

The array of quark and lepton masses and mixing angles, including
CP violation, remains a mystery. Its resolution may involve a
spontaneously broken, non-abelian flavor symmetry, a feature
common to many approaches. For example, an $SU(3)$ family
symmetry~\cite{su3,su3other} can be appended onto the standard
model or its supersymmetric extension. The flavor symmetry could
be even larger than $SU(3)$, incorporating additional, heavy
generations, a typical feature of extended technicolor models. If
the flavor symmetry is gauged, the interactions with the (broken)
gauge sector may be responsible for the observed features of the
quark and lepton mass matrices.

CP-violation is one of these features. Of all the terms in the
standard-model (SM) (dimension-4) Lagrangian which could violate
CP, those that appear in connection with flavor-changing currents
contain an $O(1)$ CP-violating phase (the CKM phase $\delta$). The
phase associated with flavor-diagonal processes is extremely small
($\bar\theta$ in the strong CP problem). This suggests that the
measured CP violation could be associated with a flavor gauge
group, and that spontaneous breaking of the latter may lead to the
breaking of discrete symmetries such as CP, and ultimately control
CP-violation in flavor changing
processes~\cite{nelsonbarr,ETC2,lane}.

In this paper, we study more generally the spontaneous breaking of discrete
symmetries in spontaneously broken gauge theories. Our main focus is on an
$SU(3)$ gauge theory, where the $SU(3)$ may be thought of as flavor group.
We consider only the gauge sector, dispensing with the standard-model
fields. Because the vacuum structure of such a theory is closely related to
the $8 \times 8$ (real symmetric) gauge-boson mass matrix, we find it
natural to embed the $SU(3)$ gauge theory in a larger model with an $SO(8)$
global symmetry. We then ask whether the resultant effective low energy
theory can, by itself, describe the spontaneous breaking of the discrete
symmetry.

To set the stage, we first examine a simpler example in which an $SO(2)$
gauge theory is embedded in a model with a global $SO(3)$ symmetry. Although
this embedding is not as compelling as $SU(3)$ in $SO(8)$, the model shares
many features with the $SO(8)$ model, including discrete symmetries, and its
vacuum structure can be more easily and thoroughly studied. After describing
this study, we examine the $SU(3)$ theory. In each case, since we are
interested in a limited range of energies, we neglect RG running of the
gauge couplings.

In Section II, we describe the $SO(2)$ gauge theory embedded in $SO(3)$.
With the $SO(3)$ broken spontaneously and with the gauge coupling
sufficiently weak, the effective low-energy theory involves a single massive
gauge field along with two massive scalars. They are psuedo Nambu-Goldstone
bosons (PNGB's). We identify the discrete symmetry of the model, and analyze
the symmetry structure of the vacuum by studying a scalar potential in terms
of the PNGB fields. A vacuum phase diagram emerges with the resultant
discrete symmetry depending on the relative strength of different terms. We
note that the phases can be physically distinguished through certain
selection rules.

In Section III, we turn to the $SU(3)$ gauge theory. We describe
its $SO(8)$ embedding and the discrete symmetry of the model. A
complete spontaneous breaking of the $SO(8)$ symmetry leads to 8
Goldstone fields and 20 additional scalar fields which are massive
due to the explicit breaking of $SO(8)$ to $SU(3)$ by the gauge
interaction. With $g$ sufficiently weak, the 20 scalars are
PNGB's, with masses small relative to the breaking scale. The
resultant low energy effective theory contains the PNGB's and the
8 massive gauge fields. We describe the operators of this theory,
and study the vacuum structure in terms of the PNGB fields to
determine under what conditions the discrete symmetry is broken
spontaneously. With 20 PNGB fields, the analysis is more intricate
than in the $SO(2)$ case. We conclude again, however, that the
spontaneous breaking of the discrete symmetries can be driven by
the low energy effective theory depending on the relative size of
certain low-energy parameters.

We summarize in Section IV, and discuss open questions.

\section{Analysis of an $SO(2)$ Gauge Theory, Embedded in $SO(3)$}

The three generators for $SO(3)$ may be taken to be \beq (T_i)_{jk} = -i
\epsilon_{ijk}. \label{generators}\eeq We choose $T_3$ as the generator of
the gauged $SO(2)$. We employ a set of scalar fields transforming as the
symmetric traceless rank-2 tensor, here the $5$ representation of $SO(3)$,
denoted $\Sigma_{ab}$, $a,b=1,2,3$. We assume that some underlying dynamics
at a scale $\Lambda$ spontaneously breaks the global $SO(3)$ completely,
producing a VEV for $\Sigma_{ab}$.

 To describe physics below the scale $\Lambda$, we freeze out two of the five degrees of
freedom in $\Sigma$, leaving the three Goldstone degrees of freedom. This
can be done via the two nonlinear constraints  \beq \Tr \Sigma^2 = f^2/2
\label{con1} \eeq and \beq \Tr \Sigma^3 = Af^3, \label{con2} \eeq where
$\Lambda = O(4\pi f)$ and where $A$ is a dimensionless parameter with unit
magnitude. The first constraint is invariant under a larger, $SO(5)$,
symmetry among the components of $\Sigma$, while the latter is $SO(3)$
symmetric.

If the $SO(2)$ gauge coupling were not present, the VEV could be
diagonalized by a general $SO(3)$ transformation to take the matrix form
    \beq
    F\,= diag(a, b, c), \label{F} \eeq where
$a+b+c = 0$. We then have  $a^2+b^2+c^2 \equiv f^2/2$. For general values $a
\neq b \neq c$, the $SO(3)$ symmetry is completely broken.

In the presence of the $SO(2)$ gauge coupling, the terms in the effective
low-energy Lagrangian quadratic in the covariant derivative are
    \beq
    \L \,=\,
    -\frac{1}{4}F_{\mu\nu}F^{\mu\nu}\,+\frac{1}{2}  \Tr
(\,D_{\mu} \Sigma \,)^2 \,+\,\frac{d}{2f^2}\Tr(\,D_{\mu} \Sigma^2 \,)^2
\,,\label{L}
    \eeq
where $D_{\mu} \Sigma^n =
\partial_{\mu} \Sigma^n \,+\,i g A_{\mu}
    \left[T_3,\Sigma^n \right]$ and $d$ is a dimensionless constant.
     That the most
    general two-derivative lagrangian can be written in terms of
    just these operators can be shown by making use of the nonlinear
    constraints (\ref{con1}) and (\ref{con2}). These allow the quantity $\Sigma^n$
    for $n\geq 3$ to be written as a linear combination of $\Sigma$,
    $\Sigma^2$ and the identity matrix. This together with a simple point
    transformation leads to the above Lagrangian. Thus, the two-derivative
    Lagrangian is characterized by the gauge coupling
    along with three additional parameters, $f$, $A$, and $d$.

To these terms must be added higher-derivative operators, as well as
operators with no derivatives describing the mass and interactions of the
PNGB's. To study this model within the framework of effective field theory,
we take the gauge coupling $g$ to be small ($g^{2}/16 \pi^{2}\ll 1$),
preserving an approximate $SO(3)$ symmetry.

Only one of the three Goldstone fields in $\Sigma$ is massless; for weak
gauge coupling the other two are PNGB's with masses of order $gf \ll \Lambda
$. To express the VEV of $\Sigma$, only an $SO(2)$ transformation generated
by $T_3$ is now available to rotate it. This freedom may be used to write
the VEV in the general form
    \beq
    \langle \Sigma \rangle \,\equiv
    \,\Sigma_0\,=\,e^{i\hat{\phi}_1T_1+i\hat{\phi}_2T_2} \, F \,
    e^{-i\hat{\phi}_1T_1-i\hat{\phi}_2T_2}\,, \label{Sigma0}
    \eeq
where $\hat{\phi}_1$ and $\hat{\phi}_2$ represent (space-time independent)
VEV's of the PNGB fields. The compact domain for these two parameters can be
determined to be
    \beq
    0\,\leq\,\hat{\phi}_1^2\,+\,\hat{\phi}_2^2 < \pi^2. \label{domain}
\eeq It can be seen that all points along the boundary
$\hat{\phi}_1^2\,+\,\hat{\phi}_2^2\ = \pi^2$ are all equivalent to the
origin up to an $SO(2)$ gauge transformation generated by $T_3$.

A convenient form for $\Sigma$ is
    \beq
    \Sigma = e^{i\phi_1T_1+i\phi_2T_2+i\phi_3T_3} \,
     \Sigma_0 \, e^{-i\phi_1T_1-i\phi_2T_2-i\phi_3T_3}\,,\label{Sigma}
    \eeq
    where the $\phi_a$ are the quantum fields, defined to have vanishing
VEV's. Using this expression and keeping only terms in $\L$ quadratic in the
quantum fields, we have \beqs
    \L_{2}
\,=\,&&-\frac{1}{4}F_{\mu\nu}F^{\mu\nu}\,+\,\frac{1}{2}\partial\phi_a\,
{K}_{ab}\,\partial\phi_b\, \\ \nonumber
    &&\,+\,g{W}_a\partial_{\mu}\phi_{a}A^{\mu}+\frac{1}{2}A_{\mu}
M^{2}A^{\mu}\,,
     \eeqs
with \beqs
     M^2({\hat\phi})&=&-g^2\
     \Tr\left[T_3,\Sigma_0\right]
     \left[T_3,\Sigma_0\right] +\, ....
\,, \label{M} \eeqs and with similar expressions for
$K({\hat\phi})_{ab}$ and ${W}_a$. The dots represent the
contribution of the third term in the Lagrangian (\ref{L}). From
the explicit expression (\ref{Sigma0}) for $\Sigma_0$, it can be
seen that this term gives the same dependence on the VEV's of PNGB
fields as the term shown, along with an additive constant.

One can introduce canonically normalized scalar fields $\varphi_a$ with
$a=1,2,3$, by diagonalizing the scalar kinetic term with a non-orthogonal
transformation of the quantum fields $\phi_a = Q_{ab} \varphi_b$ such that
$Q^{T}KQ = 1$, where $Q$ is a $3 \times 3$ constant matrix with inverse-mass
dimension. Identifying  $\varphi_3$ with the component that mixes with the
gauge boson, we have
    \beqs
    \L_2 \,=\,&&-\frac{1}{4}F_{\mu\nu}F^{\mu\nu}\,+\,\frac{1}{2}\partial
    \varphi_a\partial\varphi_a\, \label{L2}\\ \nonumber
    &&\,+\,g
W_aQ_{a3}\partial_{\mu}\varphi_3A^{\mu}+\frac{1}{2}A_{\mu}M^2A^{\mu}\, \eeqs
Thus $\varphi_1$ and $\varphi_2$ are the two PNGB's.

\subsection{Discrete Symmetries}

If the $SO(2)$ gauge coupling were not present, the symmetry of
the nonlinear Lagrangian (\ref{L}) would be $SO(3)$ along with a
discrete $Z_2$ symmetry which can be taken to be
$\Sigma\,\rightarrow\,C_{1}\,\Sigma\,C_{1}^{-1}$, where \beq C_{1}
= diag(-1,+1,+1). \label{C1} \eeq The $SO(3)$-breaking vacuum,
described by the diagonal $F$ matrix (\ref{F}), would have two
independent discrete symmetries which can be taken to be $C_{1}$
and \beq C_{2} = diag(+1,-1,+1). \label{C2} \eeq Other discrete
transformations can be formed by composing these two along with
$C_{-}= diag(-1,-1,-1)$, which is the identity transformation on
$\Sigma$. Thus the discrete $Z_2$ symmetry of the Lagrangian would
remain unbroken in the vacuum.

In the presence of the $SO(2)$ gauge coupling, the Lagrangian is invariant
under the discrete transformation \beqs
\Sigma\,&\rightarrow&\,C_{1}\,\Sigma\,C_{1}^{-1}\,
\\ \nonumber
A_{\mu}\,&\rightarrow&\,-A_{\mu}\,. \label{C}\eeqs Other discrete symmetries
are equivalent to this one together with a global $SO(2)$ transformation.

The possible discrete symmetries of the vacuum state, described by
$\Sigma_0$ (\ref{Sigma0}), depend on the values of $\hat{\phi}_1$
and $\hat{\phi}_2$. If both vanish, then $\Sigma_0 = F$, and the
vacuum has two discrete symmetries which, as in the absence of the
gauge interaction, can be taken to be $C_{1}$ and $C_{2}$. The
same is true for other values of $\hat{\phi}_1$ and $\hat{\phi}_2$
that amount to only a permutation of the entries of $F$. For a
generic value of $\hat{\phi}_1$ but with $\hat{\phi}_2 = 0$,
$\Sigma_0$ is invariant under the $C_1$ transformation. Similarly,
with $\hat{\phi}_1 = 0$ but generic $\hat{\phi}_2$, the vacuum
symmetry is $C_2$. For generic values of both $\hat{\phi}_1$ and
$\hat{\phi}_2$, $C_1$ and $C_2$ are broken, and the vacuum has no
symmetry, continuous or discrete.

\subsection{Additional Operators}

   We append to the effective Lagrangian
(\ref{L}) additional local operators  consistent with its symmetries. Since
we are interested in the contribution of these operators to the classical
potential, and since the VEV of $A_{\mu}$ must vanish, we restrict attention
to operators involving only $\Sigma$ (\ref{Sigma}), with no covariant
derivatives. These will naturally be generated by unknown high energy
physics beyond the cutoff $\Lambda$, and also in the loop expansion of the
Lagrangian (\ref{L}). They require the explicit breaking of the global
$SO(3)$, and with the assumption that the only such breaking, whether above
or below $\Lambda$, arises from the gauge interaction, they must enter with
powers of $g$.

We begin with operators including two factors of $T_3$, which are
generated at order $g^2$ from the Lagrangian (\ref{L}). By making
use of the nonlinear constraints (\ref{con1}) and (\ref{con2}),
they can all be written in terms of \beqs O_2 &=&
\Tr\left[T_3,\Sigma] [ T_3,\Sigma\right]\,\label{O1}\,, \eeqs
together with $ \Tr\left[T_3,\Sigma] [ T_3,\Sigma^2 \right]$ and $
\Tr\left[T_3,\Sigma^2] [ T_3,\Sigma^2 \right]$. All are $SO(2)
\times C_1$ invariant. From the explicit form of $\Sigma$
(\ref{Sigma}), it can be seen that the second two operators give
the same dependence on the PNGB fields as $O_2$ up to an additive
constant. We don't consider them further. Since $O_2$ enters at
order $g^2$ in the low energy effective theory (\ref{L}), we take
this to be the case in general. In the low energy theory, it
arises in Landau gauge from a single gauge-boson loop, with a
coefficient quadratically sensitive to the cutoff $\Lambda$.

\subsection{Vacuum Alignment in the Simplest Case}

Replacing $\Sigma$ by its vacuum value $\Sigma_0$, the operator
$O_2$ gives the potential \beqs V_2 &=& -C
g^{2}\left(\Lambda^{2}/16 \pi^{2}\right)
\Tr\left[T_3,\Sigma_{0}\right] \left[T_3,\Sigma_{0}\right]\, ,
\label{V} \eeqs where $\Sigma_0$ is given by Eq.  (\ref{Sigma0}),
and $C$ is taken to be of unit magnitude and positive, as
suggested by the low energy calculation. In terms of these
parameters and the elements of $F$ (\ref{F}), the explicit form of
the potential is \beqs V_2 &=& C g^{2}\left(\Lambda^{2}/8
\pi^{2}\right)
\left((a-b)^2 \,+\,\right. \label{V2} \\
&&3\left.\frac{a(a+2b)\hat{\phi}_1^2+b(b+2a)\hat{\phi}_2^2}
{\hat{\phi}_1^2+\hat{\phi}_2^2}
\sin^2\sqrt{\hat{\phi}_1^2+\hat{\phi}_2^2}\right)\nonumber
 \eeqs Recall that the domain is given by
Eq.~(\ref{domain}). In addition to $SO(2) \times C_1$, this
potential is invariant under reflection about the circle
$\hat{\phi}_1^2\,+\,\hat{\phi}_2^2 = \pi^2/4$.

 Minimization
of $V_2$ provides a simple example of vacuum alignment. Any
ordering of the parameters $a$, $b$, and $c$ ($= -a-b$) in $V_2$
may be adopted at the outset. It is convenient, however, to order
them such that $(a-b)^2$ is the smallest squared difference, that
is, $(a-b)^2 \leq (a-c)^2, (b-c)^2$. This corresponds to taking
$a$ and $b$ to be the entries with the same sign. It can then be
seen that $V_2$ is minimized by the choice $\hat{\phi}_1 =
\hat{\phi}_2 =0$. Since the gauge boson mass $M^{2}(\hat\phi)$ is
proportional to $V_2$ up to an additive constant, minimizing $V_2$
makes the gauge boson mass as small as possible. Other ordering
choices would lead to the same result since the $\hat{\phi}_i$
would then take on different vacuum values, either $0$ and
$\pm\pi/2$ or $\pm\pi/2$ and $0$, that simply permute the $F$
entries $a$, $b$, and $c$ in $\Sigma_0$. All these values for
$\hat{\phi}_1$ and $\hat{\phi}_2$ are equivalent, and leave the
vacuum invariant under the two discrete symmetries $C_1$ and
$C_2$.

We note that if any two of the $F$ entries $a$, $b$, and $c$ are
degenerate, the spontaneous breaking of the $SO(3)$ leaves a
residual $SO(2)$ symmetry group, which aligns with the gauged
$SO(2)$ subgroup, yielding a massless gauge boson. Again, the
vacuum is invariant under the discrete symmetries $C_1$ and $C_2$.

\subsection{Higher-Order Operators}

To make possible a pattern of spontaneous discrete symmetry
breaking in this model, additional potential terms must be
included along with $V_2$ (\ref{V}). These come from operators
with more than two $T_3$ factors, hence arising at higher orders
in $g^2$. Higher powers of the operator $O_2$ are of no interest,
since they enter with higher powers of $g^2$ and are therefore
always suppressed for small gauge coupling. There are other
operators entering at order $g^4$ and above, however, which are
typically small compared to the O($g^2$) operators, but have a
different dependence on the parameters of $F$. For certain choices
of these parameters, the O($g^2$) operators are suppressed, making
these new operators at least comparable. Among them, one is
bilinear in $\Sigma$, and can be taken to be \beqs O_4 &=&
\Tr[T_3,[T_3,\Sigma]] [T_3,[T_3,\Sigma]]\,. \label{O2} \eeqs It is
directly related to the four derivative operator $\Tr (D^2
\Sigma)^2$ which must be added to the Lagrangian ($\ref{L}$); by
using this four-derivative term and integrating out the gauge
boson fields, $O_4$ is generated.

Replacing $\Sigma$ by $\Sigma_0$, we denote the corresponding classical
potential by $V_4$, and use it as a representative of potential terms with
four factors of $T_3$. We then study the effective potential
\beqs V &=& V_2 + V_4 \,,\nonumber\,\\
&=&\frac{-Cg^2\Lambda^2}{16\pi^2}
\left\{\frac{}{}\Tr[T_3,\Sigma_0][T_3,\Sigma_0]\right.\,
\nonumber\,\\
&&+ \left.\frac{}{}B \,\Tr[T_3,[T_3,\Sigma_0]]
[T_3,[T_3,\Sigma_0]]\right\}, \eeqs where $B = O(g^2/16\pi^2)$.

To analyze this potential, we introduce the notation \beqs F&=&
\frac{f}{\sqrt{3}}{\rm
diag}\left(\cos(\alpha-\frac{\pi}{3}),\cos(\alpha+\frac{\pi}{3}),-\cos(\alpha)\right).
\eeqs The ordering convention $(a-b)^2$ $\leq$ $(b-c)^2$,
$(a-c)^2$ then reduces to \beq
-\frac{\pi}{6}\,\leq\,\alpha\,\leq\,\frac{\pi}{6}\,.
\label{alpharange}\eeq (The interval $[5\pi/6,7\pi/6]$ is
equivalent, up to a change of sign of $f$ which does not affect
the potential.)

The symmetries of $ V = V_2 + V_4$ are the same as those of $V_2$:
$SO(2)$, $C_1$, and a reflection symmetry about the circle
$\hat{\phi}_1^2\,+\,\hat{\phi}_2^2 = \pi^2/4$. These symmetries
allow us to restrict our study to the region \beqs
\left\{\begin{array}{ccc} \hat\phi_i&\geq&0\cr
\sqrt{\hat\phi_1^2+\hat\phi_2^2}&\leq&\pi/2
\end{array}\right.
\eeqs

Analysis of $V$ as a function of $B$ and $\alpha$ reveals a
variety of phases with respect to the discrete symmetries, in the
range $B\leq 0$. This is the range suggested by an estimate from
low-energy physics cutoff at $\Lambda$. The phases are described
by Fig. 1. We do not write the explicit form of the potential, but
simply note that for $ C> 0$, it has a global minimum at either
the origin or along the $\hat\phi_1=0$ or $\hat\phi_2=0$ axes,
depending on the location in the $\alpha$, $B$ plane.

On the right side of Fig. 1 ($0 \leq \alpha \leq\,\pi/6$), the
vacuum solution is $\hat{\phi}_1\,=\,0$, but $\hat\phi_2$ depends
on the parameters $B$, $\alpha$. To describe the shape of the
boundary curve there, we note that when $\hat{\phi}_1\,=\,0$, the
potential $V$ takes the form
    \beqs V&=&\frac{Cg^2f^2\Lambda^2}{32\pi^2}\left\{\frac{}{}(4\,-\,16\,B)
\sin^2 \alpha\right. \nonumber
\\
&+&[\,3\cos{2\alpha}\,-\,\sqrt{3}\sin{2\alpha} \nonumber \\
&&\,+B(-9\cos{2\alpha}\,+\,7\sqrt{3}\sin{2\alpha}\,+\,6)]\sin^2{\hat{\phi}_2}
\nonumber \\
&-&3B(\cos{2\alpha}\,+\,\sqrt{3}\sin{2\alpha}\,+\,2)\sin^4{\hat{\phi}_2}\}\,,
    \eeqs
with an analogous expression for the left side.

For a potential of the general form
$k_0\,+\,k_2\sin^2{\hat{\phi}_2}\,+\,k_4\sin^4{\hat{\phi}_2}$, a
simple argument leads to the conclusion that for $k_2<0$ and
$k_4>-1/2k_2$, one has a $C_1$-violating minimum at
$\hat{\phi}_2=\pm (1/2)\arccos{((k_2+k_4)/k_4)}$. This procedure,
with a similar one for the left side of Fig. 1, leads directly to
a phase diagram in the $\alpha$, $B$ plane, with the boundary
given by the expression \beqs
B^{\ast}&=&\frac{3\cos2\alpha\pm\sqrt{3}\sin2\alpha}
{9\cos2\alpha\pm7\sqrt{3}\sin2\alpha-6} \,\label {B*}. \eeqs The
boundary is a line of second-order phase transitions.

\begin{figure}[h]
\begin{center}
\includegraphics[width=3in]{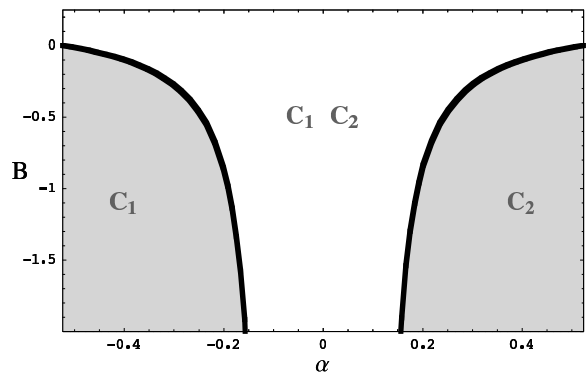}
\caption{Discrete-symmetry phases in the $\alpha$, $B$ plane. In the light
area, the minimum of the potential is at the origin. In the dark area the
minimum lies along one of the $\hat\phi_i=0$ axes. Each region is labelled
by its discrete symmetries. The boundary, given by Eq. (\ref{B*}), is a line
of second-order phase transitions \label{Fig:plot}}
\end{center}
\end{figure}

The discrete symmetry of the vacuum when $\hat\phi_1 = \hat\phi_2=0$ (the
light region of Fig.~\ref{Fig:plot}) is $C_1 \times C_2$. In the right-hand
dark region, the vacuum corresponds to the axis $\hat\phi_1=0$, and the
discrete symmetry is only $C_2$. As the right-right dark boundary line is
approached, the field $\hat\phi_2$, which is odd under $C_1$, becomes
massless. By integrating out the gauge boson and $\hat\phi_1$ fields, the
transition can be described by a Landau-Ginsberg potential in $\hat\phi_2$.
The transition along the left boundary is the same, with $\hat\phi_2
\rightarrow \hat\phi_1$ and $C_2 \rightarrow C_1$.

The physical distinction between the phases of Fig.~\ref{Fig:plot} can be
characterized in terms of the quantum fields of Eq.~(\ref{L2}). With the
operators $O_2$ and $O_4$ added to the Lagrangian (\ref{L}), and after
properly normalizing the two physical scalar fields $\varphi_i$ as in
Eq.~(\ref{L2}):

\begin{itemize}

\item
Both $\varphi_i$ fields have positive-definite masses, one of which vanishes
along the line $B=B^{\ast}$ of second-order phase transitions.

\item In the light region of Fig.~\ref{Fig:plot}, where the vacuum
is at $\hat\phi_1 = \hat\phi_2=0$, the Lagrangian has terms
containing only even powers of both quantum $\varphi_i$ fields.
The model realizes two discrete symmetries
$\varphi_i\rightarrow\pm\varphi_i$, and there is a double
selection rule. In any scattering process, the number of
$\varphi_1$ particles and $\varphi_2$ particles is conserved mod
$2$.

\item In either dark region of Fig.~\ref{Fig:plot}, the Lagrangian
is invariant under only one discrete symmetry ($C_1$ or $C_2$),
and terms that are odd in the other are present. Only one
selection rule is present. One type of particle is conserved mod
$2$, while the other is not conserved.

\end{itemize}

\subsection{Discussion}

Because $B = O(g^2/16\pi^2) \ll 1$, the parameter range of special
interest here is $\alpha \approx +\pi/6$, corresponding to the
hierarchy $|b| \ll |a|, |c|$. (The choice $\alpha \approx -\pi/6$
(the hierarchy $|a| \ll |b|, |c|$) is equivalent.)
 With only the operator $O_2$, the
effective potential is given by $V_2$, and one can see from Eq.
(\ref{V2}) that in this range the vacuum always corresponds to
$\hat{\phi}_1 = 0$. The potential becomes flat in $\hat{\phi}_2$
in the limit. Thus other operators are expected to be important in
this special case. We have analyzed $O_4$ as an example.

For $\alpha$ close to $+ \pi/6$, the potential $V = V_2 + V_4$
produces a second-order phase transition for small $B$, from the
phase with vacuum symmetry $C_1 \times C_2$ to one with only
$C_2$. With $C_1$ spontaneously broken, there is then no
(relative) small parameter in $V$, and the value of $\hat{\phi}_2$
is generically of order unity; the intrinsic breaking is large. On
the other hand, the manifestation of this effect through
selection-rule violating terms is suppressed by a small "mixing
angle" - specifically $|b/a| \approx |b/c|$.

Thus, with the small $g$ required for a consistent effective field theory,
there is a range of parameter space which allows a set of different
discrete-symmetry phases. Furthermore, the model is predictive, meaning that
the effective low energy theory can be described by a small number of
parameters ($f$ and $\alpha$, the gauge coupling $g$, and the potential
parameters). The measurement of the masses of the three physical particles
and a set of $C_i$-even couplings can determine all of these parameters. One
can then predict which discrete-symmetry phase is realized.

\section{An $SU(3)$ Gauge Theory, Embedded in $SO(8)$}

The analysis of discrete symmetries in this more realistic gauge theory is
more intricate than in the $SO(2)$ example. While the discussion here is
less complete, the features that have emerged so far are similar in many
respects to $SO(2)$.

We take the $SO(8)$ model to be described by a set of scalar fields
(composite or fundamental) transforming according to the $35$ representation
- a symmetric, traceless, rank-2 tensor denoted by $S_{ab}$,
$a,b=1,\cdots,8$. We assume the global symmetry to be completely
spontaneously broken by some unspecified underlying dynamics producing a
vacuum expectation value (VEV) for $S_{ab}$. An effective low-energy
description of this theory, valid up to the $SO(8)$ symmetry breaking scale
$\Lambda = O(4\pi f)$ and containing only the 28 Goldstone fields from the
$35$, can be constructed by freezing out the 7 massive components of $S$.
This corresponds to imposing a set of $7$ constraints on $S$, for example
$\Tr S^2 = f^2/2$.

We gauge an $SU(3)$ subgroup of $SO(8)$. A simple way to embed
$SU(3)$ in $SO(8)$ is to identify the 8 generators $(t_a)_{bc} =
-i f_{abc}$ of $SU(3)$ in the adjoint representation, and complete
this to an orthonormal basis for $SO(8)$, by adding twenty $8
\times 8$ matrices $s_b$, $b=9,\cdots,28$. This corresponds to
choosing the $SU(3)$ generators to form a maximal subalgebra of
$SO(8)$. We denote the complete generator basis as
$x_c=(t_a,s_b)$. In terms of $SU(3)$, the $28$ decomposes into $8
+ 10 + \bar{10}$. Since the $SO(8)$ is completely spontaneously
broken, so too is the $SU(3)$, and the $8$ in this decomposition
is the set of Goldstone bosons which are eaten to produce the
longitudinal component of the gauge bosons.

We write \beq \label{Eq:eff} \L \,=\, -\frac{1}{2}\Tr
F_{\mu\nu}F^{\mu\nu}\,+\frac{1}{2} \Tr(D_{\mu} S)^2 \,+\,\cdots\,, \eeq
where \beq D^{\mu} S\,=\,\partial^{\mu} S \,+\,i g \left[A^{\mu},S\right].
\eeq The gauge field $A^{\mu}_{bc}=A^{\mu}_{a} (t_a)_{bc}$ is in the $SU(3)$
adjoint representation. The dots in Eq.~(\ref{Eq:eff}) represent additional
operators involving $S$ and the gauge fields, that preserve the gauged
$SU(3)$ and other symmetries of $\L$. There are terms quadratic in the
covariant derivative $D^{\mu}$, terms involving higher derivatives, and the
zero-derivative terms of special interest here.

We parametrize the VEV $\langle S \rangle$ by introducing the notation
$\hat\phi=\hat\phi_{c}x_c$, where the $\hat\phi_c$ are $28$ real parameters
(the VEV's of the Goldstone fields), and writing
     \beqs
     \langle S \rangle \equiv
S_0&=&e^{i\hat\phi}\,F\,e^{-i\hat\phi}\,. \label{S0rep} \eeqs Here, $F=
diag(f_a)$ where the $f_a$ are eight real mass-parameters with $\sum f_a =
0$.

The quantum field $S$ can then be expressed in the nonlinear form \beq S =
e^{i\phi}\,S_{0} \,e^{-i\phi}\,, \label{Srep} \eeq with $\phi=\phi_cx_c$,
where the $28$ $\phi_c$ are the Goldstone fields, defined to have vanishing
vacuum expectation values. Since $SU(3)$ is a maximal subgroup of $SO(8)$,
there are no global continuous residual symmetries, and hence all 20
components of $\phi$ in the $SO(8)/SU(3)$ coset acquire a radiatively
generated mass and a non-trivial potential, due to the gauge interactions.
The gauge coupling is taken to be weak enough ($g^2 / 16 \pi^2 \ll 1$) so
that the scalar masses and the gauge boson masses, both of order $gf$ are
small compared to the cutoff $\Lambda$. The scalars are PNGB's.

The $SO(8)$ embedding is natural in the sense that it describes, for general
values of the $28$ $\hat\phi$ fields , the most general ($SU(3)$-breaking)
mass matrix for the $A$ fields, a real, symmetric, $8 \times 8$ matrix. The
corresponding term in $\L$, bilinear in $A$, is \beqs \label{Lm} \L_m&=&
\frac{1}{2}A^{\mu}_{a}\M^2(\hat\phi)_{ac}A_{\mu c}\,,  \eeqs where \beqs
\label{Eq:mass} \M^2(\hat\phi)_{ac}&=&-g^2\ \Tr\left\{\left[t_a,S_0\right]
\left[t_c,S_0\right]\right\}\, + .....  \,, \eeqs where the dots represent
the contribution of the additional two-derivative operators in the
Lagrangian (\ref{Eq:eff}).

The dependence of $\M^2(\hat\phi)$ on the fields $\hat\phi_a$ with $a =
1...8$ may be removed by an $SU(3)$ transformation, leaving it depending on
only the seven independent parameters in $F$ and the $20$ PNGB's in the
$SO(8)/SU(3)$ coset.

\subsection{Discrete Symmetries}

In addition to the gauged $SU(3)$, the Lagrangian is invariant under the
action of a discrete symmetry $C$, which can be identified by noting that
$SU(3)$ admits one maximal subgroup with a symmetric coset. It is an $SO(3)$
associated with the generators $t_2$, $t_5$ and $t_7$. Thus, within the
$SU(3)$, the action of $C$ is uniquely defined (up to an $SU(3)$
transformation) as: \beqs
A^{\mu}_{a}&\rightarrow&C_{ab} A^{\mu}_{b}\,,\label{Eq:C8}\\
C &=& {\rm diag}\{-1,1,-1,-1,1,-1,1,-1\}\,. \eeqs This is the canonical
definition of charge-conjugation. In terms of the matrix field
$A^{\mu}_{ab}=A^{\mu}_{c}(t_c)_{ab}$, the transformation Eq.~(\ref{Eq:C8})
becomes \beq A^{\mu}\,\rightarrow\,C\,A^{\mu}\,C^{-1}\,. \eeq

For the full $SO(8)$ theory, one can deduce from Eq.~(\ref{Eq:C8}) the
transformation properties under $C$ of all the representations. In
particular, the rank-two symmetric, traceless tensor field $S$ transforms as
     \beq
     \label{Eq:C3}
     S\,\rightarrow\,C\,S\,C^{-1}\,.
     \eeq
This is an outer-automorphism of $SO(8)$, leaving the Lagrangian
(\ref{Eq:eff}) invariant. As with $A$, $S$ is not an irreducible
representation of $C$, but decomposes into a $C$-even and $C$-odd part.
Among the 28 $\phi$ fields in $S$, the eight Goldstone fields transform just
as Eq.~(\ref{Eq:C8}), while of the remaining $20$, $10$ are $C$-odd and $10$
are $C$-even.

\subsection{Additional Operators}

Among the $SU(3) \times C$-invariant operators to be added to the Lagrangian
(\ref{Eq:eff}), are those that lead to vacuum-determining potential terms.
Since the VEV of $A_{\mu}$ must vanish, these involve only the field $S$
(\ref{Srep}). In each of these operators and the associated potential terms,
the $SU(3)$ invariance may be used to eliminate dependence on the $8$
Goldstone fields. The potential terms then depend on only the $20$ PNGB
fields $\hat\phi_{9} ..... \hat\phi_{28}$ in the coset $SO(8)/ SU(3)$. The
full potential describes a landscape in this $20$-dimensional space.

We begin at first order in $g^2$, and therefore with two factors of $t_a$.
One such operator is bilinear in $S$, and can be taken to be \beq \O_{2}
\equiv \Tr [t_{a},S][t_{a},S]. \eeq It is generated at order $g^2$ with
quadratic dependence on the cutoff $\Lambda$. (That this is the only
potential-generating operator bilinear in $S$ can be seen by noting that
$S$, which transforms as a $35$ under $SO(8)$, decomposes into $ 8 + 27$
under its maximal $SU(3)$ subgroup. There are therefore two independent
bilinears, corresponding to $8 \times 8 \rightarrow 1$ and $27 \times 27
\rightarrow 1$. But one is the $SO(8)$ invariant $ \Tr S^2 = f_{a}f_{a}
\equiv f^{2}/2$, which is independent of the $\phi_a$ fields.) With $S$
replaced by its vacuum value $S_0$, $\O_{2}$ leads to the potential \beq
{\cal V}_2 \,=\,- \C \frac{3g^2\Lambda^2
}{32\pi^2}\,\Tr\left\{\left[t_a,S_0\right] \left[t_a,S_0\right]\right\}\,,
\label{calV1} \eeq where $S_0$ is given by Eq. (\ref{S0rep}) and $\C$ is of
unit magnitude. The contribution to $\C$ coming from the low energy theory
cut off at $\Lambda$ is positive. We take this to be the case in general.

There are additional operators involving higher powers of $S$ but still
bilinear in $t_a$. By making use of the expression (\ref{Srep}), it can be
seen that, in analogy to the $SO(3)$ model, they have the same dependence on
the $\phi$ fields as $\O_2$, up to an additive constant. They lead to
potential terms that can be interpreted as simply redefining the parameters
of ${\cal V}_2$, up to an additive constant. Thus ${\cal V}_2$ can be taken
to describe the vacuum structure to order $g^2$.

\subsection{Minimization of ${\cal V}_2$}

Since $F$ is diagonal, it can be shown that all but one of the
$20$ derivatives of ${\cal V}_2$ vanish at the origin $\hat\phi_a
= 0$. The exception is the derivative along $\hat\phi_9$
corresponding to the generator $s_9$, which has non-vanishing
entries in only the `$3,8$' and `$8,3$' positions. To discuss the
stable minima, it is convenient to adopt the ordering convention
\beqs \label{order} f_8\,< \,f_4\,<\,f_5 < \, f_1\,<\,f_2\,<
\,f_6\,<\,f_7\,<\,f_3\,. \eeqs Then the potential has a stable
minimum for vanishing values of all $\hat\phi_a$ for $a = 11 ...
28$. With these choices, it is minimized in the $\hat\phi_9$
direction at a value depending on the elements of $F$. For the
case $3f_1 + 3f_2 + f_3 + f_8 > 0$, it is at \beq
\langle\hat\phi_9\rangle=-\frac{1}{\sqrt{6}}\,\arctan\frac{\sqrt{3}
(f_4+f_5-f_6-f_7)}{(3f_1+3f_2+f_3+f_8)}\,. \label{phihat9} \eeq
For the case $3f_1 + 3f_2 + f_3 + f_8 < 0$, the minimum is at
$\pi/\sqrt{6}$ plus this value.

At this minimum in all directions except $\hat\phi_{10}$, the potential is
flat in $\hat\phi_{10}$. This is the direction associated with the generator
$s_{10}$, defined such that the set $(t_3,t_8,s_9,s_{10})$ forms a basis of
the Cartan subalgebra of $SO(8)$.

The local minimum in all directions except $\hat\phi_{10}$ is not unique.
Inspection shows that there are $96=2^4\times 3!$ physically equivalent
local minima. They are all flat in $\hat\phi_{10}$, and they correspond to
values of the other $\hat\phi_a$ that simply reorder the elements of $F$ in
$S_0$, exchanging $(f_1,f_2)$, $(f_4,f_5)$, $(f_6,f_7)$ and $(f_3,f_8)$ or
permuting the pairs $(f_1,f_2)$, $(f_4,f_5)$ and $(f_6,f_7)$. All of these
physically equivalent minima are $C$-conserving; they correspond to
$\hat\phi_a$ values such that $[C,S_0] = 0$. This is evident with
$\hat\phi_a = 0$ ($a = 11, ... 28$) and $\hat\phi_{9}$ given by Eq.
(\ref{phihat9}), since $\hat\phi_9$ is a $C$ conserving direction ($[C,t_9]
= 0$). With $f$ setting the scale for each of the 19 massive $\hat\phi$'s,
the mass of each of them is of order $gf$, the same as the gauge-boson mass.
The flat direction $\hat\phi_{10}$ is, however, not $C$ conserving; thus the
determination of the $C$ symmetry of the vacuum awaits further analysis of
this direction.

The flatness in the $\hat\phi_{10}$ direction is not an accident.
It can be understood in terms of the symmetries of the system. In
any of the limits $f_1\rightarrow f_2$, $f_4\rightarrow f_5$, or
$f_6\rightarrow f_7$, an $SO(2)$ subgroup remains unbroken. Thus
there will be one less PNGB in the spectrum. In each case, one can
verify explicitly that at each local minimum the corresponding
unbroken generator is a linear combination of $s_{10}$, $t_3$ and
$t_8$ . Since the latter two are the gauged generators,
$\hat\phi_3$ and $\hat\phi_8$ will not develop a potential. Thus
$\hat \phi_{10}$ is the PNGB that decouples in each case, and any
potential term for $\hat \phi_{10}$, derived from polynomial
functions of $S_0$, must be proportional to \beqs
(f_1-f_2)\,(f_4-f_5)\,(f_6-f_7)\,. \eeqs Since this expression is
cubic in the $f$'s, ${\cal V}_2$, which is bilinear in $S_0$
cannot exhibit $\hat \phi_{10}$ dependence.

Before proceeding, a cautionary remark is in order. We have noted that a
large number of physically equivalent points exist, each a local minimum in
all directions except the flat direction $\hat\phi_{10}$. A similar feature
emerged in the analysis of the $SO(3)$ model. But unlike in that model, we
have not yet explored the full domain of $\hat\phi$ values here
\cite{mathdomain}. We conjecture that, as in the $SO(3)$ case, there will be
no further local minima degenerate with, or lower than, the ones we have
identified.

\subsection{The $\hat\phi_{10}$ Direction}

To analyze the $\hat\phi_{10}$ direction, we turn to potential terms of
order $g^4$. With $g$ small, they amount to small corrections to ${\cal
V}_1$ for the massive degrees of freedom (all except $\hat\phi_{10}$). But
they also induce a potential for $\hat\phi_{10}$. We study this
$\hat\phi_{10}$ dependence by "integrating out" the other fields of the
theory, all of which have mass scale $gf$, and focusing on energies $E \ll
gf$. This amounts to setting each $\hat\phi_a$ in $S_0$, other than
$\hat\phi_{10}$, equal to its VEV computed from ${\cal V}_2$.

We have noted that, in contrast to the $SO(3)$ model, there are no
additional, $SU(3)$-invariant operators bilinear in $S$ that can
lead to potential terms. But there are a variety of local
operators generated at the $g^4$ level, and quartic in $S$, and
they do give potentials with $\hat\phi_{10}$ dependence.

One of these, arising from the quantum loop expansion of the
Lagrangian (\ref{Eq:eff}), is of special interest. It arises in
Landau gauge from a pure-gauge-boson loop, second order in the
gauge-boson mass operator (\ref{Lm}). It enters with a coefficient
proportional to $\ln \Lambda^2$, and gives a potential
proportional to \beq \Tr(\M^2(\hat\phi))^2 \equiv
\M^2(\hat\phi)_{ab}\M^2(\hat\phi)_{ba} \, ,\label{M2} \eeq where
the gauge-boson mass matrix $\M^2(\hat \phi)_{ab}$ is given by Eq.
(\ref{Eq:mass}).

The one-loop computation leading to this potential also has logarithmic IR
sensitivity, which is conveniently regulated at momenta of order $gf$ by
summing all $\M^2(\hat\phi)$ insertions to produce the one-loop
Coleman-Weinberg potential. It takes the form \beq {\cal V}_{CW}(\hat\phi) =
\frac{3}{64\pi^2}
\Tr\left[(\M^2(\hat\phi))^2\,\left(\ln\left(\frac{\M^2(\hat\phi)}{\Lambda^2}
\right)-\frac{1}{2}\right)\right]\,, \label{CW} \eeq which incorporates the
polynomial term (\ref{M2}). The ln$\Lambda^2$ represents the contribution of
unknown physics at energies above $\Lambda^2 = 16\pi^{2} f^2$, also
receiving contributions of the same order from all levels in the loop
expansion of the effective low energy theory. We note that ${\cal V}_{CW}$
contains terms of order $g^{4} \ln g^2$.

The potential ${\cal V}_{CW}(\hat\phi)$, after integrating out all fields
except $\hat\phi_{10}$, provides nontrivial dependence on $\hat\phi_{10}$.
We examine this behavior by first focusing on the piece polynomial in
$\Tr(\M^2(\hat\phi))^2$. It leads to an effective potential as a function of
$\hat\phi_{10}$ proportional to \beqs \frac{g^{4}{\cal
D}}{16\pi^2}(f_1-f_2)(f_4-f_5)(f_6-f_7)(f_3+f_8)\,\sin^2\frac{3\hat\phi_{10}}{\sqrt{2}}\,\nonumber\,,
\eeqs where ${\cal D}$ is a constant with unit magnitude.  The extrema of
this potential are at $\hat\phi_{10}=n\,\pi/(3\sqrt{2})$ where $n$ is an
integer. Each minimum corresponds to a mass of order $g^2f / 4\pi \ll gf$,
and each leads to an $S_0$ which, up to an $SU(3)$ transformation, commutes
with $C$. Thus, whatever the sign of ${\cal D}$, this piece leads to a
$C$-conserving vacuum.

The same is true of the term proportional to $\Tr
(\M^2(\hat\phi))^{2} \ln\M^2(\hat\phi)$. This can be seen by first
noting that the gauge boson mass matrix $\M^2(\hat\phi)$ has a $2
\times 2$ block-diagonal form when, as is the case here, only
$\hat\phi_{9}$ and $\hat\phi_{10}$ are non-zero. Furthermore, only
three of these blocks depend explicitly on $\hat\phi_{10}$. The
eigenvalues of $\M^2(\hat\phi)$ are then \beqs
m^{2\,i}_{\pm}&=&a_1^i\,\pm\sqrt{a_2^{i\,2}\,+\,a_3^i\,t}\,,\\\nonumber
\eeqs where $t=\sin^23\hat\phi_{10}/\sqrt{2}$, $i=1\cdots 4$. The
$a_k^i$ are functions of the $f_a$ parameters, with, in
particular, $a_3^4=0$. Positivity and reality of the mass
eigenvalues implies that $a^{i\,2}_2+a^i_3>0$, and that $a_1^i\geq
\sqrt{a_2^{i\,2}\,+\,a_3^i\,t}$. One then finds (up to an additive
constant): \beqs {\cal
V}_{CW}(t)\,=\,\frac{3}{64\pi^2}\,\sum_{i=1}^3\,
\left((m^{2\,i}_{\pm})^2\left(\ln\frac{m^{2\,i}_{\pm}}{\Lambda^2}-\frac{1}{2}
\right)\right)\,. \eeqs

Evaluating the second derivative of this expression with respect
to $t$, we have \beqs {\cal
V}^{\prime\prime}_{CW}(t)\,=\,\frac{3}{64\pi^2}\,\sum_{i=1}^3\,\left
(\frac{a^i_3}{y^ia^i_1}\right)^2
(1+\frac{1}{2y^i}\ln\frac{1-y^i}{1+y^i})\,, \eeqs with
$y^i=\sqrt{a_2^{i\,2}\,+\,a_3^i\,t}/a_1^i$. In the allowed range
$0\leq y^i\leq 1$, the $y^i$ dependent function is negative
definite. Therefore ${\cal V}^{\prime\prime}_{CW}(t)$ is negative
definite for any value $0\leq t \leq1$, and hence the potential
${\cal V}_{CW}(t)$ has a minimum either at $t=0$ or $t=1$, where
$C$ is unbroken.

In addition to this potential term, there are several others arising at
order $g^4$ and $g^{4} \ln g^2$. We have checked that they, too, lead to a
vacuum value $S_0$ that is $C$ conserving. These details will be described
in a future paper.

\subsection{Higher Order Terms}

The key ingredient to induce spontaneous breaking of $C$ is the
presence in the potential of terms proportional to $\sin^{2n}
3\hat\phi_{10}/\sqrt{2}$, with more than one value of $n >0$. This
is analogous to the behavior found in the $SO(2)$ gauge theory. We
have noted that up to $g^4$ order, only $n=1$ terms appear, and
hence we must look to higher orders. Just as in the $SO(2)$
theory, special values of the symmetry-breaking parameters (here
the $f_a$) can suppress the $\hat\phi_{10}$ dependence at order
$g^4$, in which case terms higher order in $g$ can become
comparable in strength and induce spontaneous breaking.

Operators leading to a potential with a sum of terms with different values
of $n$ first appear at order $g^6$. They include at least six factors of the
quantum field $S$. When such operators are used to generate potential terms
through the replacement $S \rightarrow S_0$, and when all the fields
$\hat\phi_a$ are integrated out except the very light $\hat\phi_{10}$, the
potential includes terms such as $\sin^43\hat\phi_{10}/\sqrt{2}$ in addition
to ($\sin^23\hat\phi_{10}/\sqrt{2}$) behavior found at lower orders.

The interplay between these terms can lead to spontaneous breaking of $C$
just as the interplay of $V_2$ and $V_4$ did in the $SO(2)$ gauge theory. In
analogy to that example, there are values of the $SO(8)$ symmetry breaking
parameters (the $f_a$'s) that naturally suppress the $\hat\phi_{10}$
dependence of the terms lower order in $g$, allowing the higher-order terms
to compete and induce transitions to a vacuum state with broken $C$. Phase
diagrams analogous to Fig. 1 can be constructed and used to describe the
possible phases. The details of this study will be presented in a future
paper. We conclude our discussion here with three comments reminiscent of
those made in the case of the gauged $SO(2)$ model:

\begin{itemize}

\item
The values of the $f_a$'s that lead to the necessary suppression to allow
$C$ breaking are hierarchical.

\item The effective-theory framework is predictive with respect to
$C$ breaking. That is, if the ($C$-invariant) parameters of the
effective Lagrangian below the scale $\Lambda$ are measured, then
by making use of the phase diagram in these parameters, one can
predict whether $C$ is spontaneously broken. We expect the
intrinsic breaking to be of order unity as in the $SO(2)$ gauge
theory.

\item The spontaneous breaking of $C$ has physical consequences.
As in the $SO(2)$ gauge theory, there is a simple selection rule
if $C$ is unbroken: the $\phi_{10}$ particles are conserved mod 2.
If $C$ is spontaneously broken, however, odd powers are allowed
and there is no conservation of the $\phi_{10}$ particles. More
interestingly in the present case, a $C$-violating value of
$\hat\phi_{10}$ also enters the gauge boson mass matrix ${\cal
M}^{2}(\hat\phi)$. This has immediate consequences if the $SU(3)$
gauge theory is used as a flavor gauge theory coupled to the three
families of quarks. With differing up- and down-type $SU(3)$
assignments, the parameter $\hat\phi_{10}$ becomes a measurable
phase in the quark mass matrix.

\end{itemize}

\section{Conclusions}

We have analyzed the spontaneous breaking of discrete symmetries in the
effective low-energy description of a spontaneously broken gauge theory. We
have focused on an $SU(3)$ gauge theory, as well as a simpler $SO(2)$
example. To describe the breaking, we embedded these gauge theories in
models with larger global symmetry groups, $SO(8)$ and $SO(3)$ respectively.
The motivation for this is that whatever the high energy physics producing
the spontaneous breaking of the gauge group, it is likely to possess a
larger global symmetry than the gauged one. Hence PNGB's are expected to
arise, with masses controlled by the gauge coupling. Examples are (extended)
technicolor and multi-Higgs models of flavor.

We constructed the effective theory below the breaking scale $\Lambda$ of
the continuous symmetries, taking the gauge coupling small enough so that
the particles of the theory are light compared to $\Lambda$. The particle
content consists of the PNGB's produced by the spontaneous breaking of the
approximate global continuous symmetries, and the (massive) gauge bosons.
Small gauge coupling can be directly relevant to real-world gauged flavor
symmetry providing that forces other than the gauge interaction are at least
partly responsible for the breaking of the symmetry.

We studied whether the potential for the PNGB's, generated from
the operators of the low-energy theory, can trigger the
spontaneous breaking of the discrete symmetries. We found that in
both models, potentials arising at leading orders in $g$ are
minimized at discrete-symmetry preserving values of the PNGB
VEV's. In the $SU(3)$ case, this happens at order $g^2$ for all
PNGB fields except one ($\hat\phi_{10}$). For the latter, it
happens at order $g^{4}$ (and $g^{4}\ln g^2$).  But operators
generated at higher orders in the gauge coupling induce potentials
that have a different functional dependence on the PNGB VEV's, one
that can naturally lead to breaking.

For generic values of the parameters of the low energy theory, the dominance
of the lower-order terms in $g$ does not allow the higher order terms to
trigger discrete-symmetry breaking. But this breaking becomes possible if
the parameters of the effective low energy theory are such as to suppress
the PNGB-field dependence of the lower order terms. As we discussed in
detail for the $SO(2)$ gauge theory, this requires a hierarchical set of
scales in the effective theory.

There are lines of second order phase transitions in the parameter space,
and the values of the symmetry-breaking order parameter in the broken phases
is not typically small. Different phases, in which different symmetries are
preserved, are characterized by observable properties, such as couplings
that violate certain selection rules. The models are predictive, in the
sense that the knowledge of the (discrete-symmetry preserving) parameters of
the low energy theory allows one to determine in which phase the theory
lies, and compute the magnitude of symmetry-breaking interactions.

The motivation for this study was the idea that measured
CP-violation could naturally arise from the mechanism responsible
for the generation of the flavor structure of the standard model,
more specifically from the spontaneous breaking of a gauged flavor
symmetry. Our conclusion is that the spontaneous breaking of
discrete symmetries can arise purely from the gauge sector of such
theories and that it is generically expected to be intrinsically
large. This phenomenon can be analyzed within the framework of
low-energy effective field theory, without knowledge of the
details of the underlying theory and gauge-symmetry breaking,
providing that the gauge coupling is sufficiently weak. The key
necessary ingredient for discrete-symmetry breaking is the
presence of hierarchies (small mixing angles) in the low-energy
effective theory, allowing for otherwise suppressed operators to
induce the spontaneous breaking.

\acknowledgements

We thank Kenneth Lane, Nicholas Read, Robert Shrock, and Witold
Skiba for helpful conversations. This research was partially
supported by a Department of Energy grant DE-FG02-92ER-40704.

\end{document}